\documentclass[amsmath,amssymb,groupedaddress,superscriptaddress,lengthcheck,aps,prl] {revtex4-1}
\usepackage{gensymb}
\usepackage{graphics}
\usepackage{graphicx}   
\usepackage{bm}
\usepackage[latin1]{inputenc}
\usepackage{textcomp}
\usepackage{float}

\begin{document}

\title{Gapped Excitations of unconventional FQHE states in the Second Landau Level}

\author{U. Wurstbauer}
\email{wurstbauer@wsi.tum.de}
	\affiliation{Walter Schottky Institut and Physik-Department, Technische Universit{\"a}t M{\"u}nchen, 85748 Garching, Germany}
          \affiliation{Nanosystems Initiative Munich (NIM), Schellingstr. 4, 80799 M{\"u}nchen, Germany }
\author{A. L. Levy}	
	\affiliation{Department of Physics, Columbia University, New York, NY 10027, USA}
\author{A. Pinczuk}
	\affiliation{Department of Physics, Columbia University, New York, NY 10027, USA}
	\affiliation{Department of Applied Physics and Applied Mathematics, Columbia University, NY, New York 10027, USA}
\author{K. W. West}
	\affiliation{Department of Electrical Engineering, Princeton University, Princeton, NJ 08544, USA}
\author{L. N. Pfeiffer}
	\affiliation{Department of Electrical Engineering, Princeton University, Princeton, NJ 08544, USA}
	\author{M. J. Manfra}
\affiliation{Department of Physics and Astronomy, Birck Nanotechnology Center, Purdue University, West Lafayette, IN, USA}
 \affiliation{School of Materials Engineering, Birck Nanotechnology Center, Purdue University, West Lafayette, IN, USA}
	 \affiliation{School of Electrical and Computer Engineering Birck Nanotechnology Center Purdue University, West Lafayette, IN, USA}
\author{G. C. Gardner}
	 \affiliation{School of Materials Engineering, Birck Nanotechnology Center, Purdue University, West Lafayette, IN, USA}
\author{J. D. Watson}
	 \affiliation{Department of Physics and Astronomy, Birck Nanotechnology Center, Purdue University, West Lafayette, IN, USA}

\date{\today}
\begin{abstract}
We report the observation of low-lying collective charge and spin excitations in the second Landau level at $\nu = 2+1/3$ and also for the very fragile states at $\nu = 2+2/5$, $2+3/8$ in inelastic light scattering experiments. These modes exhibit a clear dependence on filling factor and temperature substantiating the unique access to the characteristic neutral excitation spectra of the incompressible FQHE states. A detailed mode analysis reveals low energy modes at around 70 $\mu$eV and a sharp mode slightly below the Zeeman energy interpreted as gap and spin wave excitation, respectively. The lowest energy collective charge excitation spectrum at $\nu=2+1/3$ exhibits significant similarities and a universal scaling of the energies with its cousin state in the lowest Landau level at $\nu=1/3$ suggesting similar underlying physics. The observed excitation spectra facilitate to distinguish between theoretical descriptions of the nature of those FQHE states.  A striking polarization dependence in light scattering is discussed in the framework of anisotropic electron phases that allow for the stabilization of nematic FQHE states.
\end{abstract}


\maketitle
Ultra-pure two-dimensional electron systems subjected to high perpendicular magnetic fields form diverse quantum ground states that are driven by  strong Coulomb interactions between electrons. In a partially populated N=0 Landau level (LL) Fractional Quantum Hall effect (FQHE) states are interpreted as weakly interacting quasiparticles of electrons with even numbers of vortices of the many-body wavefunction attached to the electrons (known as composite fermions (CFs)) \cite{Jain2007}. The quantum phases in higher LLs (N$>$1) are governed by different interaction physics \cite{Lilly1999, Du1999}. The second Landau Level (SLL) with N=1 is special since odd-denominator FQHE states as well as unconventional FQHE states such as the enigmatic even-denominator states at $\nu=5/2$ \cite{Willett1987} and $\nu = 7/2$ compete with other ground states. Competing phases manifest in transport experiments in an anisotropic longitudinal resistance and as reentrant integer quantum Hall effect (RIQHE)  \cite{Xia2004,Deng2012,Friess2014}. For $\nu = 2+1/3=7/3$ a large anisotropy in the resistance and a robust FQHE state are in coexistence indicating that the FQHE can be stabilized in absence of full rotational invariance \cite{Xia2011,Mull2011,Qiu2012,Liu2013}. It has been proposed that transport anisotropies in the SLL can be explained in terms of nematic electron liquid, a compressible metallic phase that is expected to exhibit strong signatures in polarized light scattering experiments due to unequal longitudinal and transverse susceptibilities $\chi_{L}$ and $\chi_{T}$ \cite{Fradkin2000}.
\par
The nature of both, the more conventional as well as unconventional FQHE states in the SLL are not yet well-known. Similarly, their low-lying collective excitations spectra that are unique fingerprints of each state are neither theoretically well understood nor experimentally observed. The excitations of the FQHE state at $\nu=2+1/3$, the cousin of the most robust state at $\nu=1/3$, are predicted as composite fermions dressed with an exciton cloud \cite{Balram2013}. The authors state that the 1/3 and 2+1/3 could be determined by the same physics and the exciton screening impacts the 2+1/3 state only quantitatively without changing its nature \cite{Balram2013}. Besides the much studied FQHE state at $\nu=5/2$, interpreted as a p-wave paired state of composite fermions supporting non-Abelian excitations \cite{Moore1991}, the state at $2+2/5$ is envisioned as an exotic parafermionic state \cite{Bonderson2012}. It has been suggested that the weaker $2+2/5$ and $2+3/8$ FQHE states exhibit even greater potential than the 5/2 state to serve as model system for fault-tolerant quantum computation \cite{Bonderson2008,Bonderson2012}. 
\par
Collective charge and spin excitations of phases in the N=0 lowest LL (LLL) are accessed by resonant inelastic light scattering (RILS) methods \cite{Pinczuk1993, Kang2001, Dujovne2003, Hirjibehedin2005, Dujovne2005}, and quantitative comparisons of the measured low-lying excitation spectra with theory provide in-depth understanding of the physics driving the emergence of those quantum states \cite{Rhone2011a, Wurstbauer2011}. Interpretations of measured low-lying excitations in the N=1 LL from theoretical formulations of the underlying quantum phases would offer further insights into interaction physics in the SLL.
\par
\begin{figure}[htpb]
\centering
\includegraphics[width=0.5\textwidth]{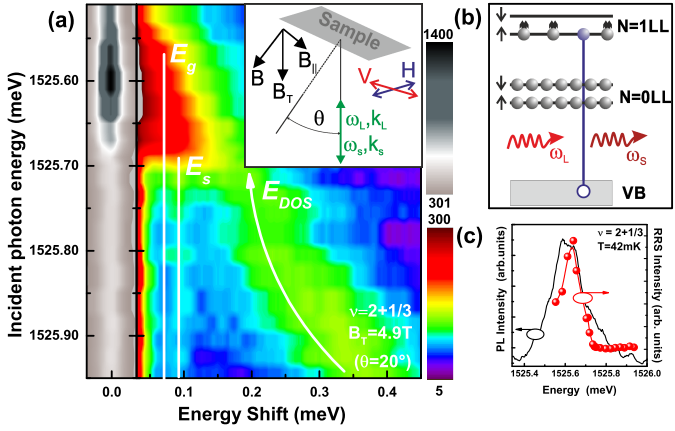}
\caption{(color online) (a) Color plot of RRS (grey scale) and RILS intensities (color scale) for photon energies close to the optical emission from the N = 1 LL at $\nu=2+1/3$ as a function of the exciting light energy $\omega_{L}$ measured at temperature T=42mK. The marked dependence of line shapes on $\omega_{L}$ is due to a strong outgoing resonance in RILS \cite{note1}. Three modes are seen at energies $E_{s}$, $E_{DOS}$ and $E_{g}$. Inset: Light scattering geometry and the magnetic field direction. The red and blue arrows denote the linear polarization of photons. (b) Energy level scheme for incoming and outgoing photon energies ($\omega_{L(S)}$) resulting in resonant enhancement in RRS and RILS spectra. The energy is close to the optical transition between valence band and spin-up branch of the N=1 LL. (c) Emission spectrum from the N=1 LL (black line) and related RRS intensities (red dots) obtained from the spectra shown in (a).}
\label{fig:fig1}
\end{figure}
\par
In this letter we report RILS observations of a remarkable filling factor dependence of low-lying excitations of the partially populated SLL in the range $5/2 > \nu > 2+1/5$.  RILS spectra are interpreted in terms of density of states of large wave vector modes that are activated by residual disorder. The modes exhibit a marked filling factor dependence and are only well developed for filling factors that are known from transport to form incompressible, albeit weak FQHE states such as $\nu$ = 2+2/5, 2+3/8 and 2+1/3 \cite{Xia2004, Pan2008, Kumar2010}. Energy gaps identified in these measurements are well below 0.1~meV (about  1K). These observations suggest that the FQHE states seen in transport in the filling factor range $2+2/5 > \nu > 2+1/5$\ have well-defined low-lying excitation modes that manifest the underlying interaction physics.
\par
We find that RILS spectra at the filing factors of these weak FQHE states typically display three distinct modes with intensity that is resonantly enhanced as shown in Fig. \ref{fig:fig1}(a) for $\nu = 2+1/3$, the most robust odd-denominator FQHE state in the SLL. In Fig. \ref{fig:fig1}(a) there is a band with a maximum that shifts with $\omega_{L}$ and occurs in the energy range from 0.15~meV $< E_{DOS} <$ 0.35~meV. In addition there is a broad mode centered at $E_{g} \approx 0.08$~meV, and a weak sharp mode at $E_{s} \approx 0.1$~meV. The modes $E_{DOS}$ and $E_{g}$ are interpreted as spin conserving excitations of the quantum fluid. The $E_ {s}$ mode is assigned to a low-lying excitation with spin reversal \cite{Wurstbauer2013}. Comprehensive mode analysis at 2+1/3 uncovers that the mode labeled $E_{g}$ can be decomposed into two modes as will be described below. A quantitative comparison with the calculated as well as measured excitation spectrum of the 1/3 state \cite{Rhone2011a} indicates astonishing agreement between the lowest energy mode dispersion for the 2+1/3 and 1/3 state by introducing an universal scaling by a factor of $0.15 \pm 0.01$. This finding suggest a similar underlying physics between the $\nu = 2+1/3$ state in the SLL and the $\nu = 1/3$ state in the LLL.
\par 
\begin{figure}[htbt]
\centering
\includegraphics[width=0.5\textwidth]{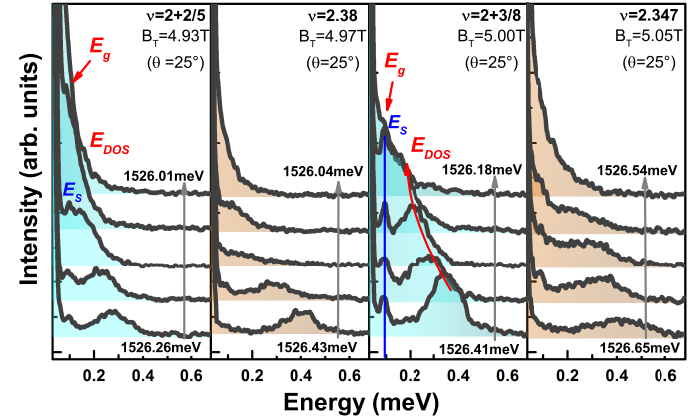}
\caption{Filling factor dependence of RILS spectra for $2+2/5 \geq \nu \geq 2.347$. The spectra are shifted vertically for clarity. All observed, resonantly enhanced modes exhibit a striking filling factor dependence, drawing attention to the filling factors $\nu = 2+2/5$ and $\nu = 2+3/8$ that are known from transport to be incompressible FQHE states. (\textit{(H,V), T = 42~mK, $\theta=25\degree$}).} 
\label{fig:fig2}
\end{figure}
\par
By tuning the filling factor away from the magic filling fractions the charge modes $E_{g}$ and $E_{DOS}$ almost disappear and the spin-mode $E_{s}$ is significantly reduced as shown in Fig. \ref{fig:fig2} and in the SOM \cite{note6}. Surprisingly, even the unconventional FQHE states at $\nu=2+2/5$ and $2+3/8$, known to be fragile in activated transport \cite{Xia2004, Pan2008, Kumar2010}, exhibit well defined low energy modes in RILS spectra measured at the elevated temperature of T = 42~mK. The distinct dependence on filling factor as well as on temperature of the three low-lying modes  in Figures \ref{fig:fig2} and SOM \cite{note4,note6} substantiate the link to incompressible quantum states. Interestingly, we observe a pronounced dependence of the RILS modes on photon polarization, which is most remarkable for the lowest energy mode $E_{g}$ (see Fig. \ref{fig:fig5}). This observation is linked to the occurrence of nematic liquids induced by the application of finite in-plane magnetic fields \cite{Fradkin2000,Xia2011, Qiu2012, Liu2013}. The experimentally explored low-lying excitation spectra of the puzzling $2+1/3$, $2+2/5$ and $2+3/8$ FQHE states pave the way to distinguish between different scenarios about their nature provided by theory.
\par
\begin{figure}[htpb]
\centering
\includegraphics[width=0.45\textwidth]{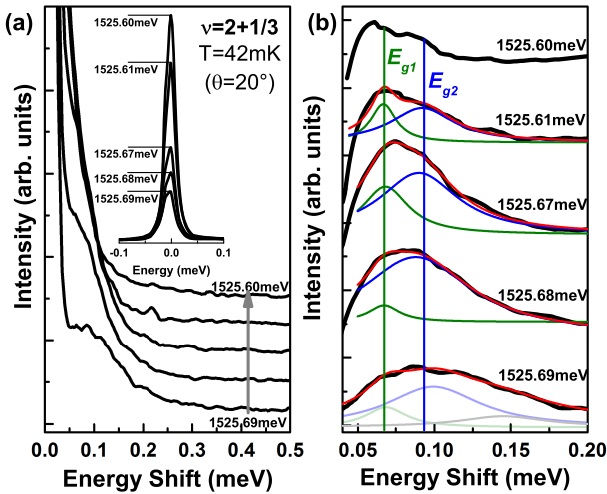}
\caption{Mode analysis of the broad resonantly enhanced low energy mode $E_{g}$ for $\nu = 2 + 1/3$ (\textit{(H,V), T = 42mK, $\theta = 20\degree$}). (a) Individual RILS spectra excited with relevant photon energies $\omega_{L}$, vertically shifted for clarity. The inset displays the same spectra around zero energy highlighting the RRS contribution. (b) Spectra shown in (a) after subtraction of the RRS intensities. A mode analysis with 2 Lorentzian uncovers two resonantly enhanced low energy modes at  $E_{g1} \approx 67\mu$eV = $7.9 \times 10^{-3}$ E$_{c}$ and $E_{g2} \approx 90\mu$eV = 1.06 $\times 10^{-2}$ E$_{c}$, respectively, with E$_{c}$ = e$^{2}$/$\epsilon l$ and $l$ the mangnetic length.}
\label{fig:fig3}
\end{figure}

The ultra-clean 2D electron system is confined in a 30 nm wide symmetrically doped single GaAs/AlGaAs quantum well structure. The charge carrier density and mobility determined from transport at T = 300 mK are $2.9\times10^{11} \textnormal{cm}^{-2}$ and $23.9\times10^{6}$ $\textnormal{cm}^2/\textnormal{Vs}$, respectively. The measurements have been done in a $^{3}\textnormal{He}/^{4}\textnormal{He}$ dilution refrigerator with a 16 T magnet and bottom windows for optical access. The RILS and resonant Rayleigh scattering (RRS) spectra are excited by a Ti:Sapphire laser at a power below $10^{-4} \rm W/\rm cm^{2}$. The energy of the light $\omega_{L}$ is tuned to be close to the optical emission from the N = 1 LL as sketched in Fig.\ref{fig:fig1} (b) to achieve resonant enhancement \cite{Rhone2011b,Wurstbauer2013}. Emission from the N=1 LL and RRS spectra  are displayed in Fig. \ref{fig:fig1}(c). The used backscattering geometry is sketched in inset of  Fig. \ref{fig:fig1}(a). The sample is tilted at an angle $\theta = 20\degree$ or $\theta = 25\degree$ in two different cool-downs, respectively, to allow the transfer of a finite momentum $k=|\vec{k_{L}}-\vec{k_{S}}|=(2\omega_{L})/c)\sin \theta$, where $\vec{k_{L(s)}}$ is the in-plane component of the incident (scattered) photon, $\omega_{L}$ the incoming photon energy and $c$ the speed of light. The tilt angle results in a small in-plane magnetic field component $B_{||}$  that still allows well defined FQHE states at $\nu=5/2$ and $\nu=2+1/3$ and the formation of anisotropic phases in the second LL \cite{Friess2014, Pan1999,Csathy2005,Xia2010,Xia2011}. The spectra taken at different cool-downs with slightly different tilt angles of $\theta = 20\degree$ and $\theta = 25\degree$ are apparently looking very similar (compare e.g. spectra displayed in Figs.\ref{fig:fig1}, \ref{fig:fig3} and SOM \cite{note6}). The filling factor as a function magnetic field is precisely determined from the maximum of the spin-wave intensity, an excitation occurring at the bare Zeeman energy $E_{Z}$, for $\nu = 3$ \cite{Wurstbauer2013,note2}.
\par
The polarization of incoming and scattered light is denoted with V and H as sketched in inset of Fig. \ref{fig:fig1}(a) and described in detail in the SOM \cite{note5}. 
\par
\begin{figure}[ht]
\centering
\includegraphics[width=0.45\textwidth]{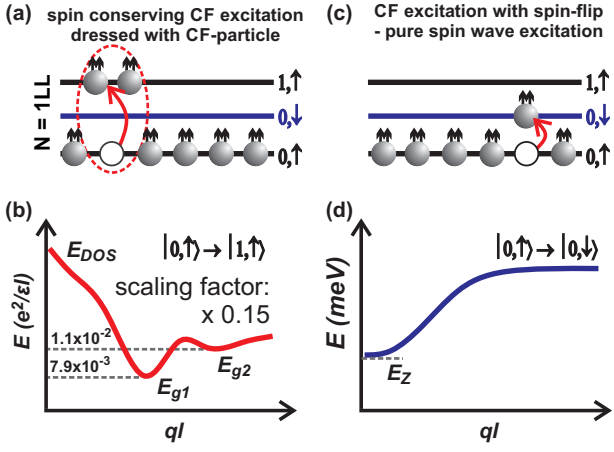}
\caption{Spin-split $\Lambda$-levels, that are Landau levels of CFs, within the N=1 LL and pictorial description of CF charge excitation consisting of a CF quasiparticle 'dressed' with a spin-conserving CF excitation \citep{Balram2013} in (a) and a spin wave excitation (c). The related wave-vector dispersion are scaled down from the calculated dispersion of $\nu=1/3$ state in the LLL in panels (b) and (d), respectively (modified from \citep{Rhone2011a} and \cite{Wurstbauer2011}).} 
\label{fig:fig4}
\end{figure}
\begin{figure}[htpb]
\centering
\includegraphics[width=0.45\textwidth]{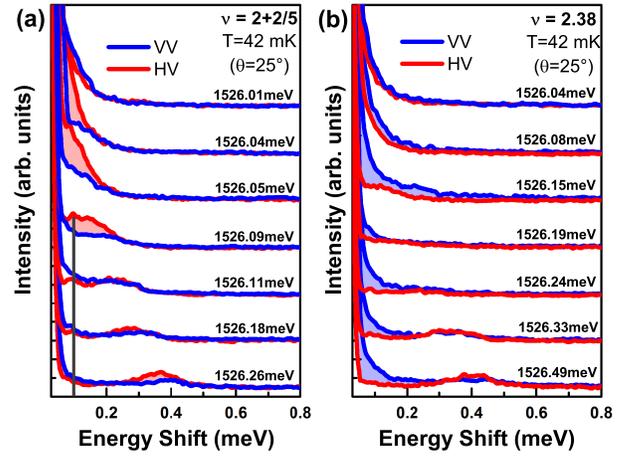}
\caption{Polarization dependent RILS spectra in (H,V) geometry (red) and (H,H), geometry (blue) for (a) $\nu$ = 2+2/5 and (b) $\nu = 2.38$, respectively. (\textit{T = 42~mK $, \theta=25\degree$}).}
\label{fig:fig5}
\end{figure}

In Fig. \ref{fig:fig1}(a) the (HV) RILS spectra display three features that are interpreted as collective excitations modes of the incompressible quantum fluid at $\nu = 2+1/3$.  Intensity maxima are assigned either to critical points in the wave vector dispersion with a high density of state (DOS), such as rotons or maxons activated by breakdown of wave vector conservation due to residual disorder \cite{Wurstbauer2011}, or to long wavelength modes with $k=q$ \cite{Pinczuk1993}. In this framework the sharp mode $E_{s}$ is interpreted as wave-vector conserving $q \rightarrow 0 $ spin wave excitation at the bare Zeeman energy E$_{Z}=\mu_{B} g B$, where $\mu_{B}$ is the Bohr magneton and $g$ the bare $g-$factor. The deviation of the measured $g$-factor $|g|$=0.36 from the value for free electrons in bulk GaAs  ($|g|$=0.44) has previously been reported \cite{Wurstbauer2013} and can be seen as precursor for breakdown of full rotational invariance \cite{Gallais2008, Drozdov2010}. The mode $E_{s}$ exhibits a less pronounced temperature dependence. The peak is slightly broadened at 250mK and  still observable at 600~mK consistent with the interpretation of $E_{S}$ as a pure spin wave mode that is broadened by increasing the temperature, but not much affected by melting of an incompressible fluid. The related excitation scheme and mode dispersion are depicted in Fig. \ref{fig:fig4}(c,d).
\par
The intense mode labeled $E_{g}$ in RILS spectra at $\nu = 2+1/3$ was investigated by subtraction of the RRS intensity as shown in Fig. \ref{fig:fig3}. The measured spectra are regarded  as superposition of the RILS signal and of a strong RRS (plotted in the inset of Fig.\ref{fig:fig3}(a) and in Fig.\ref{fig:fig1}(a)).  The two contributions to the light scattering intensities can be decomposed by subtracting a Lorentzian fit to the RRS signal centered at $E=0$~meV from the measured spectra. The subtracted spectra shown Fig. \ref{fig:fig3}(b) reveal that at low energy the RILS component of the measured spectra can be well described by two Lorentzian peaks centered at $E_{g1} \approx 67\mu$eV = $7.9 \times 10^{-3}$ E$_{c}$ and $E_{g2} \approx 90\mu$eV = 1.06 $\times 10^{-2}$ E$_{c}$, respectively, with with E$_{c}$ = e$^{2}$/$\epsilon l$ and $l$ the magnetic length. 
\par
We interpret the two modes $E_{g1(2)}$ as lowest energy collective spin-preserving excitations of the $2+1/3$ FQHE state. The lowest neutral excitation is theoretically predicted by Balram $\textit{et al.}$ to be built of CF particles or holes dressed with a spin-conserving charge excitation \cite{Balram2013} as sketched for the CF particle in Fig. \ref{fig:fig4}(a). Within the framework of breakdown of wave vector conservation, the modes are expected to occur at critical points in the wave vector dispersion \cite{Wurstbauer2011,Rhone2011a} and are assigned to a roton minimum $\delta_{R}$ at finite $q$ and to the large momentum limit $\Delta_{\infty}$ at $q \rightarrow \infty$ as depicted in Fig. \ref{fig:fig4}(b,d). Similarly, the RILS intensity at $E_{DOS}$ with a low-energy onset at around 0.15~meV $\approx 1.76 \times 10^{-2}$ E$_{c}$ is attributed to the mode $\Delta_{0}$ at the long wavelength limit $q \rightarrow 0$. The assignment of the modes has been done in analogue to the wave-vector dispersion of the well understood cousin state at $\nu = 1/3$ in the LLL. It is striking that a quantitative comparison between the mode energies at 2+1/3 and 1/3 results in an universal scaling factor of about $0.15 \pm 0.01$ suggesting similar underlying physics of the two states in the SLL and its cousin state in the LLL as considered to be possible by theory \citep{Balram2013}. The scaling factor may be due to larger extension of the wave function of the quasiparticle in the second Landau level and the fact that the quasiparticle excitations are dressed by an exciton cloud \citep{Balram2013}.
The interpretation of the modes as collective excitation of the incompressible quantum state is strongly supported by their temperature dependence \cite{note4}. The RILS intensity of modes at $E_{g}$ are already significantly reduced by rising the temperature from 42~mK over 100~mK to 250~mK, and are absent in the spectra at 600~mK. Similarly, the mode labeled $E_{DOS}$ gets broadened and greatly reduced in intensity by increasing the temperature \cite{note4}. A quantitative mode analysis of the $E_{g}$ band at 2+2/5 and 2+3/8 is demanding and uncovers only one mode at $E_{g}$ that is centered below 75$\mu$eV. A more exact analysis is hindered by a combination of ultra-low energies, weakness of the modes and smaller range of resonance enhancement. Temperature dependent measurements reveal that the modes, particularly $E_{g}$, are already significantly reduced by increasing the temperature from 42~mK to 65~mK and are further weakened by raising the temperature to 100~mK. The strong temperature dependence of the modes underlines the fragility of quantum fluids at these filling factors. 
\par
The polarization dependence of modes observed in RILS is exemplified by the results at filling factor $\nu=2+2/5$ shown in Fig. \ref{fig:fig5}. While it is known that excitations with spin reversal are more intense in cross-polarized scattering, in the presence of an external magnetic field RILS polarization selection rules are relaxed so that spin as well as charge modes are accessible in cross- as well as co-polarized scattering \cite{Dujovne2003}. RILS experiments are typically performed in cross-polarization to suppress parasitic light at $\omega_{L}$ that would mask the signal of RRS  and of low-energy RILS modes as shown for $\nu=2+2/5$ and $\nu=2.38$ in the SOM \cite{note3}. The spectra in Fig. \ref{fig:fig5} were obtained by careful suppression of parasitic light at $\omega_{L}$ to allow quantitative analyses of RILS and RRS spectra in (H,V) and (V,V) geometries. In these results, the lowest energy mode E$_{g}$ and, as expected, the weak mode $E_{s}$ are weak in (V,V) spectra. In addition, at $2+2/5$ the RRS signal is much stronger in (H,V) \cite{note3}. The gapped modes are absent for both, (H,V) and (V,V) scattering geometries at filling factors slightly away, by $\Delta \nu = 0.02$,  from $\nu=2+2/5$. Simultaneously, the RRS in (H,V) is significantly reduced. It is evident that in non-resonant excitation the intensity at zero energy is higher for (V,V) compared to (H,V) independent from the filling factor due to parasitic intensity at $\omega_{L}$.  Both, RRS and RILS spectra exhibit a striking polarization dependence only for filling factors linked to an incompressible FQHE state. We ascribe the polarization dependence in inelastic as well as elastic light scattering to anisotropic susceptibilities $\chi_{\parallel}$ and $\chi_{\perp}$ parallel and transverse to the in-plane component of the magnetic field $B_{||}$ as predicted by theory \cite{Fradkin2000}. The filling factor dependence further corroborates the interpretation from transport experiments that nematic FQHE states are stabilized in the SLL at $\nu=2+1/3$ \cite{Xia2011}, $\nu=5/2$ \cite{Liu2013} and  $\nu = 2+2/5$ \cite{Zhang2012}. This interpretation is consistence with the redshift of the SW energy resulting in a reduced value of the $g$-factor due to the collapse of full rotational invariance. 
\par
To summarize, gapped low energy modes have been observed in RILS for $\nu=2+2/5$, $2+3/8$ and $2+1/3$. Even for the very fragile states at $2+2/5$ and $2+3/8$ three modes are clearly observable and are interpreted as collective spin and charge modes of the FQHE states.  This interpretation is corroborated by the clear filling factor and temperature dependence of the modes. A detailed mode analysis done for the most robust state in the SLL at $\nu=2+1/3$ suggests that the excitation spectrum exhibits qualitative as well as quantitative agreement with the 1/3 state in the LLL taking an universal scaling factor into account suggesting similar underlying physics between the two FQHE states. Observations from polarization dependent RILS and RRS measurements at $\nu=2+2/5$ can be explained by an anisotropic susceptibility consistent with the existence of nematic FQHE states in the SLL in presence of an in-plane magnetic field \cite{Xia2011,Liu2013,Friess2014}. The reported results provide in-depth inside to the nature of the fragile and enigmatic FQHE states in the SLL and can facilitate to distinguish between different theoretical scenarios. 
\par
We would like to thank J. Jain and E. Fradkin for insightful comments to the manuscript and A. F. Rigosi for support during measurements. The work at Columbia is supported by the National Science Foundation Division of Materials Research (NSF DMR) under  Awards DMR-1306976 and DMR-0803445, and by the Alexander von Humboldt Foundation. The research at TUM was supported by the DFG via Nanosystems Initiative of Munich (NIM). The MBE growth and transport characterization at Princeton was supported by the Gordon and Betty Moore Foundation under Award GMBF-2719 and by the National Science Foundation, Division of Materials Research, under Award DMR-0819860. The MBE growth and transport measurements at Purdue are supported by the U.S. Department of Energy, Office of Basic Energy Sciences, Division of Materials Sciences and Engineering under Award DE-SC0006671. 
\bibliographystyle{apsrev4-1}

%

\end{document}